# Scattered Image Reconstruction at Near-infrared Based on Spatial Modulation Instability

Yuan Liao, Lin Li, Zhaolu Wang, Nan Huang, and Hongjun Liu

*Abstract*—We present a method of near-infrared image reconstruction based on spatial modulation instability in a photorefractive strontium barium niobate crystal. The conditions that lead to the formation of modulation instability at near-infrared are discussed depending on the theory of modulation instability gain. Experimental results of scattered image reconstruction at the 1064 nm wavelength show the maximum cross-correlation coefficient and cross-correlation gain are 0.57 and 2.09 respectively. This method is expected to be an aid for near-infrared imaging technologies.

## I. Introduction

WHEN optical images propagate through a haze, turbid water, or biologic tissues, light scattering and signal distortion cannot be avoided. Imaging technologies via ballistic light extraction, such as range-gating, polarization-based gating, and adaptive optics, have been widely applied to remote sensing, underwater detection, and biological imaging [1-3]. In contrast, computational imaging technologies, including the use of wavefront-shaping, optical memory effect, and correlation holography, have shown that images could also be restored from speckle patterns [4-7].

Spatial modulation instability (MI) indicates that small intensity perturbations hiding in a uniform beam could be amplified by self-focusing [8-11]. In the area of nonlinear imaging, incoherent spatial MI in photorefractive media enables an energy transfer of optical noise to underlying images (stochastic resonance) [12-17], and visible image (532 nm wavelength and white light) reconstruction has been experimentally demonstrated by strontium barium niobate (SBN) crystals [12-15]. The method works whether signals are affected by scattering or background noise. Particularly, it is effective in the cases where strong background noise and scattering noise are existing simultaneously, which are challenging for both ballistic light extraction and computational imaging.

Near-infrared imaging has been used in biomedicine, night vision, and LiDAR for decades [18-24]. In particular, light with the 1064 nm wavelength has the advantage in atmospheric propagation and falls within the detection range of a silicon-based camera [25-27]. Despite the lower electro-optic coefficient and sensitivity, photorefractivity at 780 to 1550 nm wavelengths is still promising in SBN crystals [28-31]. In this paper, we present that SBN crystals allow the near-infrared image reconstruction. Specifically, the preconditions of spatial MI forming at 1064 nm are theoretically studied; experiments on scattered image reconstruction at 1064 nm are then performed, where the transmitted image is scattered by a rotating scatterer, and the ballistic light induces MI and hence partially recovers the original image within a biased SBN crystal.

## II. Theoretical Analysis

In self-focusing media, inevitable small fluctuations at the wavefront of a uniform beam tend to grow. Whereas the grown fluctuations intensify self-focusing in turn. As a consequence, patterns caused by initial instability form at the output of media (spatial MI). Propagation of light with slowly varying amplitude $E$ in self-focusing media is described by the following parabolic wave equation [32]

$$2ik\frac{\partial E}{\partial z} + \nabla_\perp^2 E + 2n_0 \frac{\omega^2}{c^2}\Delta n E = 0. \tag{1}$$

MI gain (mm$^{-1}$) to the fluctuations within a spatial incoherent beam in photorefractive crystals with screening nonlinearity is derived from (1) and given by [9]

$$g = |k_\perp|\sqrt{\frac{\Delta n_0}{n_0}\frac{I_0/I_d}{(1+I_0/I_d)^2} - \left(\frac{k_\perp}{2k}\right)^2} - |k_\perp|\theta_0. \tag{2}$$

$k$, $\omega$, $n_0$, $\Delta n$, $g$, $k_\perp$, $\theta_0$, and $I_0$ are the wavenumber, the frequency, the initial refractive index, the index change, the MI gain coefficient, each transverse spatial frequency of perturbations, the angular spectrum width, and the optical intensity, respectively. $\theta_0 = 2\pi / kl_c$ is relevant to the spatial coherence $l_c$ and negatively correlated with gain. $\Delta n_0 = (1/2)n_0^3 \gamma_{33} E_0$ is the index change induced by the applied electric field $E_0$, where $\gamma_{33}$ is the linear electro-optic coefficient. Dark irradiance is defined as [33]

$$I_d = \frac{\beta}{s}, \tag{3}$$

here $\beta$ is the thermal emission rate and $s$ is the photoionization coefficient depending on the wavelength $\lambda$.

This work was supported in part by the National Natural Science Foundation of China under Grant 61775234 and Grant 61975232 (*Corresponding author: Hongjun Liu*).

Yuan Liao and Lin Li are with State Key Laboratory of Transient Optics and Photonics, Xi'an Institute of Optics and Precision Mechanics, Chinese Academy of Sciences, Xi'an 710119, China, and University of Chinese Academy of Sciences, Beijing 100084, China (e-mail: liaoyuan18@mails.ucas.ac.cn; lilin183@mails.ucas.ac.cn).

Zhaolu Wang and Nan Huang are with State Key Laboratory of Transient Optics and Photonics, Xi'an Institute of Optics and Precision Mechanics, Chinese Academy of Sciences, Xi'an 710119, China (e-mail: wangzhaolu@opt.ac.cn; huangnan@opt.ac.cn).

Hongjun Liu is with State Key Laboratory of Transient Optics and Photonics, Xi'an Institute of Optics and Precision Mechanics, Chinese Academy of Sciences, Xi'an 710119, China, and Collaborative Innovation Center of Extreme Optics, Shanxi University, Taiyuan 030006, China (e-mail: liuhongjun@opt.ac.cn).

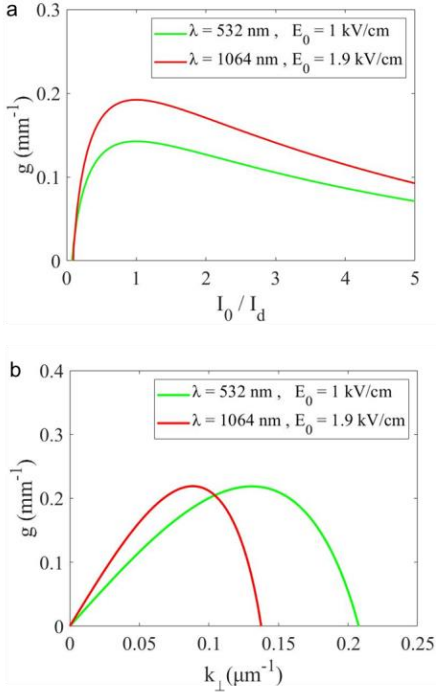

Fig. 1. (a) Gain coefficient of modulation instability vs the ratio of incident intensity to dark irradiance for different wavelengths and applied electric fields. (b) Gain coefficient of modulation instability vs transverse spatial frequency. Other parameters are fixed at: $n_0 = 2.3$ and $\gamma_{33} = 270$ pm/V at $\lambda = 532$ nm; $n_0 = 2.2$ and $\gamma_{33} = 220$ pm/V at $\lambda = 1064$ nm; $\theta_0 = 1.8$ mrad; (a) $k_\perp = 0.063$ $\mu m^{-1}$; (b) $I_0 / I_d = 1$.

MI gain coefficient as functions of normalized intensities and transverse spatial frequencies are respectively shown in Fig. 1. In SBN crystals, photoionization $s$ rapidly recedes as long as wavelengths exceed the range of visible light [$I_d$ ($\lambda = 0.5$ $\mu$m) ≈ 0.1 mW/cm$^2$; $I_d$ ($\lambda = 1$ $\mu$m) ≈ 20 W/cm$^2$] [29, 31]. Nevertheless, MI could still occur at near-infrared only if $I_0 / I_d$ remains unchanged (Fig. 1a). Also, gain reaches the peak value at $I_0 / I_d = 1$. However, MI would be unobservable at $I_0 \ll I_d$. Light with the 1064 nm wavelength experiences a weaker electro-optic effect within the crystal, so a higher electric field is required to achieve the gain as that at 532 nm (Fig. 1). The intuitive explanation for the nonlinear process of the MI-based scattering imaging method is as follows: the ballistic light (light without scattering) induces intensity redistribution of scattered light due to MI; the increasing electric field strengthens self-focusing, which leads to more accumulations of scattered light on the underlying pattern till the steady state. According to Fig. 1b, MI leads to different levels of growth for different transverse spatial frequencies $k_\perp$. Therefore, effective image reconstruction achieves if the characteristic spatial frequencies of images fall within the range including the maximum MI gain.

## III. EXPERIMENTS AND RESULTS

The experimental setup of scattered image reconstruction is shown in Fig. 2. A continuous-wave laser beam ($\lambda = 1064$ nm; maximum power, 120 mW) is focused onto a transmissive target (resolution test target, 1951 USAF). A dc voltage is applied along the crystalline c-axis of the 5 cm × 5 cm × 8 mm SBN:61 (Sr$_{0.61}$Ba$_{0.39}$Nb$_2$O$_6$) crystal and parallel to the polarization of the beam (extraordinary polarization). The image is scattered directly by a rotating scatterer before imaged onto the back surface of the crystal. An aperture stop before the imaging lens is used to collimate the signal beam. Light exiting the crystal is attenuated and then captured by a CMOS digital camera. The angular spectrum width $\theta_0$ of scattered light is controlled by the distance $d$ between the target and the scatterer. Fig. 3 shows the experimental results of the reconstructed '2' images with increasing nonlinearity. The applied electric field is adjusted within the range of $E_0 = 0$ kV/cm to 4 kV/cm. The optical intensity is fixed at $I_0 = 6.1$ W/cm$^2$. The distance between the target and the scatterer are (b–f) $d = 270$ mm and (g–k) $d = 190$ mm respectively. Image quality is estimated by the 2-D cross-correlation coefficient (0 to 1) between output images and original images and the cross-correlation gain $G_c = C_{out} / C_{in}$ represents the improvement. Fig. 4 summarizes the cross-correlation coefficient and the cross-correlation gain as a function of the applied electric field for images in Fig. 3. Image quality improvements are able to be observed. If nonlinearity could be further increased (e.g., $E_0 = 8$ kV/cm), however, spontaneous MI for higher spatial frequencies would blur the images [12, 13]. For scattered images, the degree of distortion is concerned with the intensity of ballistic light. We find experimentally that the intensity of ballistic light relies on $d$ (or $\theta_0$) and it can also be verified easily by geometrical optics. When setting the distance between the target and the scatterer from $d = 270$ mm to $d = 190$ mm, one can find that the image quality at each applied field improves with decreasing scattering noise (Figs. 3, 4). The cross-correlation coefficients for varying optical intensity $I_0$ are summarized in Fig. 5. Linear behaviors of the cross-correlation coefficient versus the applied field are observed, confirming that the intensities are less than $I_d$ at 1064 nm [31]. As the theoretical model has been predicting, the image reconstruction performances reduce with optical intensity.

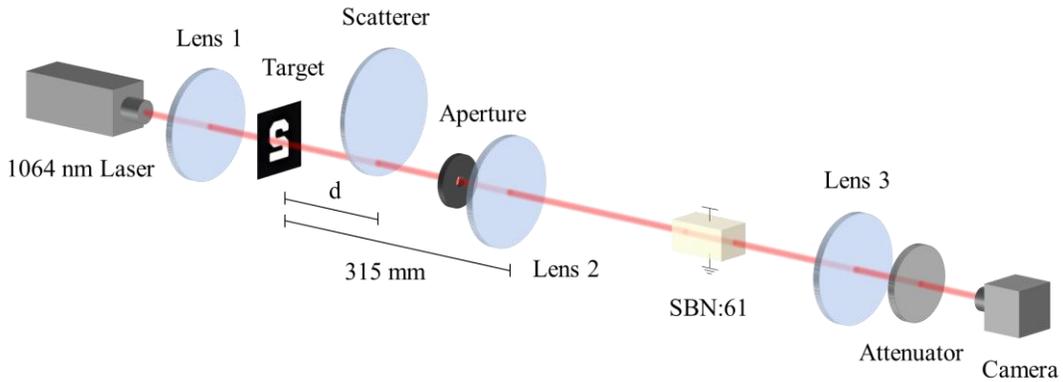

Fig. 2. Experimental setup. A 1064 nm laser beam is focused onto a transmissive target. The position of the scatterer determines the angular spectrum width of scattered light, and a biased voltage controls the nonlinear strength of the SBN crystal. Before imaged into a CMOS digital camera, light exiting the crystal is attenuated. Light propagating length in the crystal, $l = 8$ mm.

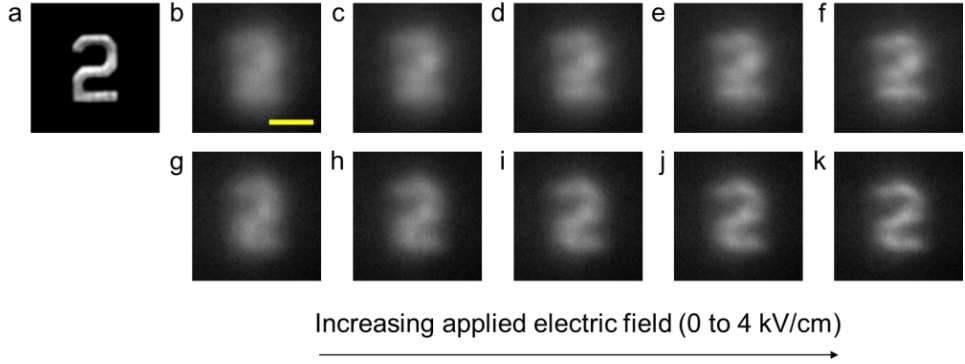

Fig. 3. Experimental results for reconstructed '2' images. Optical intensity, $I_0 = 6.1$ W/cm$^2$. (a) Original image. Scattered images at the output face with (b) $d = 270$ mm and (g) $d = 190$ mm. (c–f) and (h–k) are reconstructed from (b) and (g) respectively with applied electric fields of (c, h) 1 kV/cm, (d, i) 2 kV/cm, (e, j) 3 kV/cm, and (f, k) 4 kV/cm. Scale bar, 150 μm.

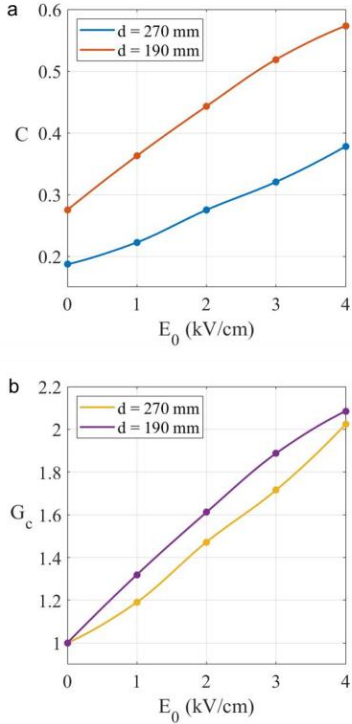

Fig. 4. (a) Cross-correlation coefficient and (b) cross-correlation gain vs applied electric field corresponding to the images in Figs. 3(b–f) and (g–k).

In electrically biased crystals, free carriers tend to drift and be captured along the electric field. As a result, the index modulation is asymmetry, and therefore the horizontal parts of '2' are reconstructed more effectively than the vertical parts under a vertically applied field (Figs. 3f, 3k). Further observations of reconstructed '1' images confirm this type of optical anisotropy. At $E_0 = 4$ kV/cm, visibility is enhanced on the top and bottom of '1' (the horizontal parts), whereas the middle is nearly changeless (Figs. 6e, 6f). In comparison, it is the opposite when the object is rotated 90° (Figs. 6c, 6d). (We rotate the object rather than the crystal as the polarization is required to be parallel to the applied field.) Naturally, higher image quality is obtained as long as we average the images reconstructed from the two varying orientations (Fig. 6b).

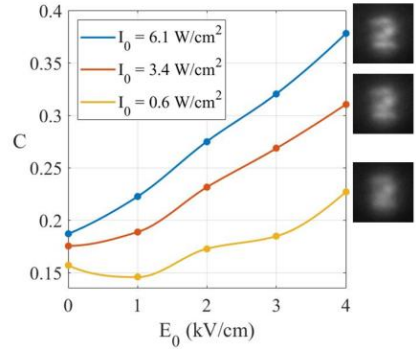

Fig. 5. Cross-correlation coefficient vs applied electric field for different values of the incident intensity. Other experimental conditions for $I_0 = 3.4$ W/cm$^2$ and 0.6 W/cm$^2$ remain the same as $I_0 = 6.1$ W/cm$^2$.

## IV. CONCLUSION

We have presented a nonlinear method for near-infrared imaging through a scatterer. Also, the image reconstruction method based on induced MI in photorefractive SBN crystals has been expanded to near-infrared imaging. The cross-correlation coefficients and the cross-correlation gains of reconstructed images with the 1064 nm wavelength reach 0.57 and 2.09 respectively. We have also demonstrated the influences of varying experimental conditions, such as electric fields, optical intensities, scattering, image characteristics, and carriers' move and capture on the image reconstruction performance. This method has potential for a variety of near-infrared imaging applications, and it can be easily extended to other near-infrared wavelengths.

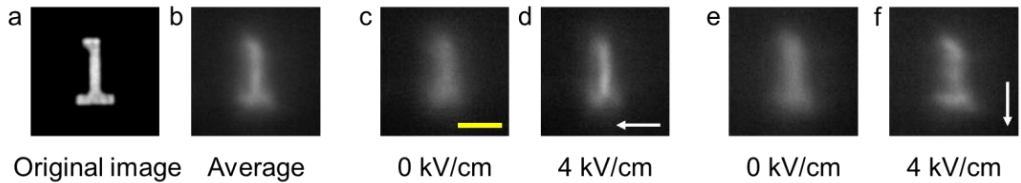

Fig. 6. Experimental observations of optical anisotropy. (a) Original '1' image. (c) Scattered image rotated 90°. (e) Scattered image without rotation. (d) and (f) are reconstructed from (c) and (e) respectively. The arrows point toward the electric field directions. (b) Average of (d) and (f). Cross-correlation coefficient of each output image, (b) 0.8046 (maximum), (c) 0.6221, (d) 0.7538, (e) 0.7186, and (f) 0.7766. Scale bar, 150 $\mu$m.